\documentclass[twocolumn,prb]{revtex4}
\usepackage{graphicx,amsmath}

\begin{document}

\title{A simple model of Coulomb disorder and screening in graphene}

\author{B. I. Shklovskii}

\affiliation{Theoretical Physics Institute, University of
Minnesota, Minneapolis, Minnesota 55455}

\date{\today}

\begin{abstract}

We suggest a simple model of disorder in graphene assuming that
there are randomly distributed positive and negative centers with
equal concentration $N/2$ in the bulk of silicon oxide substrate.
We show that at zero gate voltage such disorder creates the
two-dimensional concentrations $n_0 \sim N^{2/3}$ of electrons and
holes in the graphene sample. Electrons and holes reside in
alternating in space puddles of the size $R_0 \sim N^{-1/3}$. A
typical puddle has only one or two carriers in qualitative
agreement with the recent scanning single electron transistor
experiment.

\end{abstract}
\maketitle

Recent experiments on the gate voltage dependent 2D transport of
graphene deposited at SiO
substrate~\cite{Novoselov04,Novoselov05,Kim05} attracted almost
unprecedented theoretical attention (see, for
example~\cite{Mirlin,Sarma,Adam,Galitski,MacDonald,Aleiner}, and
references therein). Clean graphene has zero mass Dirac
relativistic spectrum near the Fermi level $(\varepsilon = 0)$,
where energies $\varepsilon$ of both electrons and holes behave as
\begin{equation}
\varepsilon = \pm \hbar v k. \label{spectrum}
\end{equation}
Here the velocity $v \simeq 10^6$ m/s and $k $ is absolute value
of electron wave vector. Density of states for such a (four times
degenerate) spectrum $\nu(\varepsilon) = 2\varepsilon /\pi
\hbar^{2}v^{2}$ vanishes at $\varepsilon = 0$. Only states with a
negative energy are filled in a clean graphene sample.

Experimental measurements of the electrical conductivity and the
Hall effect voltage strongly indicate presence of disorder. Most
of the theoretical discussion is focused around observed
phenomenon of saturation of the linear dependence of
two-dimensional conductivity $\sigma$ on the gate voltage at small
gate voltages, where the conductivity reaches the minimum value
$\sigma = \sigma_{min} = Ce^2/h$. Particular attention is
attracted by the constant $C$ in this
formula~\cite{Mirlin,Aleiner}.

In this note we concentrate on the width of the conductivity
minimum. Using the linear relation between gate voltage and
concentration of electrons we characterize this width by the
concentration $n = n_0$, where saturation starts. The hint to the
meaning of $n = n_0$ follows from the Hall effect data, well
fitted by the so-called two band model, which assumes simultaneous
presence of electrons and holes even at zero gate
voltage~\cite{Novoselov04}. It was
suggested~\cite{Katsnelson,Sarma,Adam,Galitski} that one can
understand these phenomena assuming that potential of charged
impurities moves the Dirac point up and down in different points
of space creating alternating in space electron and hole puddles.
Theoretical self-consistent calculations of the random potential
and carrier distribution~\cite{Sarma,Adam,Galitski} so far were
based on models of two-dimensional distribution of charged
impurities, which require at least two parameters, the total
two-dimensional concentration of impurities and width of the layer
near the surface they reside.

Here we would like to suggest a simple model of distribution of
charged impurities and  a theory of their nonlinear screening.
Below we assume that there is three-dimensional concentration
$N/2$ of both positive and negative centers randomly distributed
in the bulk of SiO substrate and estimate all mentioned above
quantities in terms of the single parameter of the model $N$. We
estimate the concentrations of electrons and holes $n_0$ at the
zero gate voltage (or, more exactly, in the center of minimum of
conductivity), the characteristic spatial scale $R_0$ of
nonlinearly screened potential (which gives the size of the
typical electron or hole puddle and the distance between
neighboring electron puddles) and the typical number of carriers
in a puddle in terms of $N$. For this purpose we study screening
of the random potential of 3D impurities by a graphene sample.

Nonlinear screening of the random potential of charges impurities
by electron and hole puddles was first studied for
three-dimensional totally compensated
semiconductors~\cite{ES1972,ES1986,book}. Following these works we
start from calculation of the random potential with the spatial
scale $R$. We imagine that all oxide is divided in cubes with the
edge length $R$. Each cube has average number of impurities
$NR^{3}$ and fluctuating excessive charge of either sign
$e(NR^{3})^{1/2}$. This charge creates the random potential
fluctuating from a cube to cube with the amplitude of the order of
\begin{equation}
V(R) = (e/\kappa R)(NR^{3})^{1/2} = (e/\kappa)(NR)^{1/2},
\label{VR}
\end{equation}
where $\kappa \sim 2.5$ is the effective dielectric constant at
the oxide surface. Here and everywhere below we drop numerical
coefficients making only order of magnitude estimates. All the
cubes adjacent to the surface of oxide apply this potential to the
graphene sample. At zero gate voltage (in conductivity minimum)
there are no electrons and holes without this potential. The
random potential moves $\varepsilon = 0$ point up and down and
creates electrons in its wells and holes on its hills. Let us
concentrate on electrons. Electrons screen a typical well of the
size $R$ almost completely if they can fit in this well in the
number equal to the well charge. The concentration of electrons
necessary for neutralization of the well charge is
\begin{equation}
n(R) =(NR^{3})^{1/2}/R^{2}= (N/R)^{1/2} . \label{NR}
\end{equation}
Let us assume that the wavelength of these electrons $n^{-1/2}$ is
smaller than $R$, or $nR^{2} \gg 1$, so that the Fermi energy of
them electrons can be estimated in the Thomas-Fermi approximation,
\begin{equation}
E_F(R) = \hbar v n^{1/2} = (N/R)^{1/4} . \label{EF}
\end{equation}
If $E_F(R)\ll eV(R)$ all electrons compensating the well charge
are localized inside the well close to its bottom and are
fragmented in space by the random potential of even smaller sizes.
Thus we are dealing with nonlinear screening when concentration of
electrons is strongly nonuniform.

The process of fragmentation of electron liquid continues until
$eV(R) \gg E_F(R)$. It stops at such $R=R_0$, where $E_F(R_0) =
eV(R_0)$. This gives equation for $R_0$
\begin{equation}
\hbar v (N/R_0)^{1/4} = (e^{2}/\kappa)(NR_0)^{1/2}, \label{RO}
\end{equation}
which yields $R_0 = N^{-1/3} (e^2/\kappa \hbar v)^{-4/3}$. Using
$v = 10^{8}$ cm/s we get that $e^2/\kappa \hbar v \sim 0.8$ and
$R_0 \sim N^{-1/3}$. The smallest wells of the size $R_0$ are not
fragmented and form what one may call electron puddles. A typical
puddle has one or two electrons. Groups of one of two negative
impurities at the distance $N^{-1/3}$ from graphene form hole
puddles. Condition $nR^{2} \gg 1$ used above is only marginally
correct, so that the whole calculation is only an order of
magnitude estimate.

Now we can estimate the two-dimensional concentrations of
electrons and holes. Each of them occupy half of space with the
concentration
\begin{equation}
n_0 \simeq n(R_0) \simeq N^{2/3}.  \label{NO}
\end{equation}
Eq.~(\ref{NO}) is the main result of our paper. It gives the
two-dimensional concentration of coexisting electrons and holes
$n_0$ in graphene in terms of a single parameter of our model,
three-dimensional concentration $N$ of charged impurities. It
could be easily anticipated because this is the only formula with
necessary dimensionality one can make from $N$. There are no large
or small dimensionless parameters in the system, which could play
the role of large or small coefficient in Eq.~(\ref{NO}). Indeed,
the single dimensionless parameter $e^2/\kappa \hbar v $ is close
to unity.

Above we are using the term puddle size for the smallest scale of
fragmentation of electron and hole density. Although electrons and
holes are really located in such small puddles there are
fluctuations of the electron density at larger scales $R \gg
N^{-1/3}$. At such scales the electron density follows Gaussian
fluctuations of the density of impurity charges. For example, for
domains of size $R$ the electron concentration $n(R)$ is given by
Eq.~(\ref{NR}) and is much smaller than $n_0$. Although some
people may refer to these domains also as "puddles", they are
actually small gaussian fluctuations of the density of puddles of
one or two electrons we are talking about. In other words, in the
plane of graphene large scales of the random potential $R \gg R_0$
are screened linearly, by small variations $n(R)$ around the
average concentration $n_0$ of electrons and holes situated in
puddles. One can check that at $e^2/\kappa \hbar v \sim 1 $ the
linear screening radius corresponding to the two-dimensional
concentration  $n_0$ is equal to $R_0$.

Recently both the concentration of carriers $n_0$ and the
characteristic size of puddles $R_0$, were estimated by
compressibility measurements based on the use of the scanning
single electron transistor~\cite{Yacoby}. The authors arrived at
$n_0 = 2.3~ 10^{11}$cm$^{-2}$ and $R_0 = 30$~nm. These numbers
result approximately in $n_0 R_{0}^{2} \sim 2$ electrons per
puddle in a good agreement with our estimates. Using
Eq.~(\ref{NO}) we can also estimate that the three-dimensional
concentration of charges in the silicon oxide substrate $N \sim
10^{17}$ cm$^{-3}$, what is not unreasonable.

We would like to emphasize that while working with limited
resolution authors of Ref.~\cite{Yacoby} saw  domains of
fluctuations of density of electrons with larger sizes $R \gg
R_0$, but with much smaller concentrations $n(R) \ll n_0$. They
noticed that $n(R)$ increases with the decreasing scale $R$, and
improving experimental resolution traced observed gaussian
dependence of $n(R)$ to the smallest scale cited above. Their
observations are in agreement with our picture.

In the recent paper~\cite{Aleiner} authors suggested a theory of
graphene conductivity assuming existence of macroscopic puddles
totally filled by electrons and holes and separated by narrow
$p$-$n$ junctions. In other words, they assumed that each electron
puddle has many electrons. We see that in our model macroscopic
approach to puddles is not justified. This may lead to some
numerical changes in calculation of constant $C$ in the expression
for $\sigma_{min}$ above and in the shape of the conductivity
versus gate voltage curve near the minimum, but definitely will
not change the qualitative conclusions of Ref.~\cite{Aleiner}.

Note that fragmentation of the electron density in graphene into
puddles of one or two electron is not an exclusive feature of the
model of three-dimensional distribution of impurities. Similar
results follow from random distribution of charges in a
two-dimensional plane close to
graphene~\cite{Sarma,Adam,Galitski}. Note also a difficulty with
application of the model of this paper for evaluation of the
Coulomb scattering limited mobility of electrons at large
concentrations $n \gg n_0$. Indeed at large Fermi wave vectors
$\sim n^{1/2}$ electrons are strongly scattered only by the the
Coulomb centers at the distance of the electron wavelength
$n^{-1/2} \ll N^{-1/3}$ from the graphene plane. In other words,
electrons do not "see" more distant centers. As a result the
mobility should increase proportionally to $n^{1/2}$. This
prediction is in contradiction with the observed independent on
$n$ mobility, which in turn is well explained by Coulomb
impurities residing in graphene~\cite{MacDonald}. This may be a
strong argument for the dominating role of charged centers located
in graphene in the currently studied samples. But even if this is
proven, in future charges in graphene may be neutralized or
eliminated. Then one will arrive to the case the bulk charged
impurities discussed here.

I am grateful to V. Galitski, D. Novikov and K. Novoselov for very
useful discussions.

%------------------------------------------------------------------------%

\end{document}